\documentclass[review,12pt]{elsarticle}

\usepackage{amssymb}
\usepackage{amsmath,esint}

\usepackage{lineno}
\usepackage{xcolor}

\journal{Acta Astronautica}

\begin{document}

\begin{frontmatter}



\title{Fully kinetic study of facility pressure effects on RF-source magnetic nozzles}


\author[inst1]{Raoul Andriulli}
\cortext[cor1]{Corresponding author}
\ead{raoul.andriulli@unibo.it}

\author[inst1]{Shaun Andrews}

\author[inst2]{Nabil Souhair}

\author[inst3]{Mirko Magarotto}

\author[inst1]{Fabrizio Ponti}

\affiliation[inst1]{organization={Alma Propulsion Laboratory, Department of Industrial Engineering, University of Bologna},
            addressline={via Fontanelle 40}, 
            city={Forli},
            postcode={47121}, 
            country={Italy}}

\affiliation[inst3]{organization={Department of Information Engineering (DEI), University of Padova},
            addressline={via Gradenigo 6/b}, 
            city={Padova},
            postcode={35131},
            country={Italy}}

\affiliation[inst2]{organization={LERMA Laboratory, Aerospace and Automotive Engineering School, International University of Rabat  },
            addressline={Sala al Jadida}, 
            city={Rabat},
            postcode={11100},
            country={Morocco}}

\begin{abstract}
A fully kinetic 2D axisymmetric Particle-in-Cell (PIC) model is used to examine the effects of background facility pressure on the plasma transport and propulsive efficiency of magnetic nozzles. Simulations are performed for a low-power (150 W class) cathode-less radio-frequency (RF) plasma thruster, operating with xenon, between background pressures up to $10^{-2}$ Pa and \textcolor{black}{average electron} discharge temperatures of $4-16$ eV. When the electron temperature within the near-plume region reaches 8 eV, a decisive reduction in performance occurs: at $10^{-2}$ Pa, in-plume power losses surpass 25\% of the discharge energy flux. \textcolor{black}{Given that the ionization energy for Xe is 12 eV, the 8 eV threshold indicates that a consistent percentage of electrons has energy enough to trigger ionization.} On the other hand, \textcolor{black}{when} the temperature is below \textcolor{black}{such} threshold, the primary collisions are charge-exchange and inelastic ion scattering, \textcolor{black}{and} the power loss remains less than 10\%. It is established that losses in the considered HPT are significant if the facility pressure is greater than $10^{-3}$ Pa, at absorbed powers larger than 130 W. At the nominal 150 W, this results in a 15\% thrust reduction. When facility pressure is taken into consideration over ideal vacuum simulations, numerical error is reduced to $<$30\% when compared to experimental thrust measurements at $10^{-3}$ Pa.
\end{abstract}


\begin{keyword}
Particle-in-Cell \sep cathode-less RF plasma thruster \sep facility pressure \sep magnetic nozzle \sep xenon 
\end{keyword}
\end{frontmatter}


\section{Introduction}
\label{sec:introduction}

In recent years, the increasing demand for simple and low-cost electric propulsion for small satellites has given rise to a growing interest in low-power ($<$250 W) cathode-less magnetically-enhanced plasma thrusters (MEPT) \cite{b:keidar2014}. This category primarily includes the Helicon Plasma Thruster (HPT) \cite{b:HPT} and the Electron Cyclotron Resonance Thruster (ECRT) \cite{b:ECR}. In such systems, the plasma acceleration is driven by a magnetic nozzle (MN): a divergent magneto-static field generated by a set of solenoids or permanent magnets. The MN radially confines the hot quasi-neutral partially-magnetised plasma beam and accelerates it via the conversion of thermal energy into directed axial kinetic energy, therefore enhancing thrust \cite{b:magarotto2020}.

MEPTs can operate on a wider range of propellants (e.g iodine and water) \cite{b:iepciodine}; these are often less expensive than traditional xenon because of their greater abundance, and are easier to store due to the non-requirement of pressurised tanks. \cite{b:bellomo2021,b:nakagawa}. From a systems perspective, MEPTs can also be less-complex than state-of-the-art devices (e.g. gridded ion and Hall effect thrusters (HET)) as they do not require a separate, dedicated neutralising electron source (e.g. a hollow cathode), which increases the power and propellant consumed, and is a known failure mechanism \cite{b:failure}. Due to the absence of plasma-immersed electrodes, MEPTs are highly resistant to erosion, which is often the lifetime-limiting aspect of conventional electric propulsion systems \cite{b:mazouffre}. MNs also have no physical walls, thus avoiding thermal loading and further erosion issues.
However, since the ion acceleration occurs external to the plasma source --- in a region that can be several times the thruster radius --- devices that use a MN are highly susceptible to so-called facility effects \cite{jarrige2014measurements, zhang2021}. EP thrusters interacting with the facility environment is a widely recognised problem in the field and calls into question the ability to extrapolate measured performance of such systems to their intended environment in space \cite{aerospace7090120, cai2015}.

While facility effects continue to be an on-going area of research for most forms of EP, MEPTs pose a particular challenge: experimental efforts have revealed that their response to facility effects, particularly the presence of excess neutrals due to finite pumping speed, is different from other state-of-the-art EP technologies \cite{b:minotor}.
The experiments of Vialis et al. \cite{vialis2017geometry}, with an ECRT operating on xenon at 40 W, revealed that the high-vacuum thrust efficiency of 12.5\% at 6$\times10^{-4}$ Pa would drop to 9\% at 7.2$\times 10^{-4}$ Pa, and 3.9\% at 1.3$\times 10^{-3}$ Pa facility pressure. It was found that the divergence of plasma within the plume increased, on-axis ion current density decreased by almost 60\%, and the thruster floating potential decreased by a factor of approximately 0.25 as background pressure was raised. Studies of a 100 W HPT, that were conducted at higher facility pressures of 1.9$\times 10^{-3}$ - 8.8$\times 10^{-3}$ Pa, revealed similar trends \cite{b:caruso2018}. In both cases, the changes in efficiency and divergence were counter to what is found for more established forms of EP such as HETs \cite{aerospace8030069,kerber2019,piragino2021,hargus2013,snyder2020}. These latter technologies typically improve in performance with increasing pressure. Kerber et al. \cite{kerber2019} describes how the beam, current, and mass utilization efficiencies of a system are affected by the background pressure. The study found that decreasing background pressure resulted in a decrease in mass and current efficiencies \textcolor{black}{(the latter defined as the ratio of beam to discharge currents)} of the tested HET, while the beam efficiency \textcolor{black}{(related to the divergence angle)} was only slightly affected.
Snyder \cite{snyder2020} discusses the dependence of absolute HET thrust on pressure and power. The study found that the highest power had the largest dependence on pressure, with little-to-no measurable dependence at the lowest power. However, the relative change of thrust with pressure was consistent for all powers above 1 kW, with about $2-4\%$ higher thrust at 1.3$\times10^{-3}$ Pa compared to the lowest facility pressure at each power level ($\sim4\times10^{-4}$ Pa). The arguments that have been employed for these systems to explain facility effects thus do not seem to apply to devices utilising a MN.

Several theories have previously been put forward to explain the performance detriment of MNs due to non-negligible facility pressure. These have included background neutral ingestion impacting the power balance and consequent plasma production in the source region \cite{b:caruso2018}, as well as energy losses due to the onset of secondary discharges occurring outside the thruster plume when the neutral environment is sufficiently high \cite{vialis2017geometry}.
Vialis \cite{vialis2017geometry} and Collard and Jorns \cite{b:collard2019} have shown that the presence of excess neutrals can also directly impact the dynamics of the MN itself, i.e. the ability to convert thermal energy into directed kinetic energy. Experiments demonstrated, for example, that the acceleration profile of ions is reduced at high facility pressure. This was numerically confirmed by Andrews et al. \cite{b:andrews2022}, who suggested that ion drag from charge and momentum exchange collisions in the plume substantially delays the ambipolar ion acceleration. Wachs et al. \cite{wachs2018effects}  found that the performance loss can largely be attributed to electron–neutral inelastic collisions within the plume that significantly reduce the amount of power available to accelerate ions. The same experimental campaign showed that potential drop decreased by 20\% as background pressure increased from 9$\times 10^{-5}$ Pa to 7$\times 10^{-3}$ Pa. From additional tests conducted in \cite{wachs2020background}, with a background pressure varying between 1.3$\times 10^{-4}$ Pa and 3.4$\times 10^{-3}$ Pa, Wachs calculated that, at maximum pressure, almost 40\% of the power entering the plume was consumed mainly by inelastic collisions. 
This correlates with the results of previous numerical work \cite{b:andrews2022,b:AA2022}, where it was clear that power entering the plume, in the form of both ion inertia and electron pressure, is consumed by ionisation and excitation collisions with the neutral gas plume of the thruster. Physically, these collisions remove the critical thermal energy introduced to the electrons in the source region before it can be successfully converted to ion kinetic energy. This however, only applies to high-$T_e$ (i.e.~$>$10 eV) plasma sources where inelastic collisions dominate, neglecting other hypothesised effects such as increased plasma friction, cross-field transport and early detachment from the MN. 

It is therefore critical that numerical models must be augmented to accurately describe plume dynamics in thrusters operating in finite background pressure conditions. The aim of this study is to conduct a thorough investigation of the performance loss mechanisms induced by non-negligible facility pressures across a wide range of thruster operating powers.
To this end, a recently developed numerical suite \cite{b:AA2022,andrews2023}, which consists of a 0D Global Source Model (GSM) of the plasma source, coupled to a fully kinetic Particle-in-Cell (PIC) model of the MN, is used to investigate in detail facility pressure effects on the MN performance of a 150 W class MEPT. Section 2 describes the GSM and PIC models, and the numerical setup. In section 3, the thrust and power balances in the MN region are presented to enable characterisation of the loss mechanisms. Section 4 is dedicated to the presentation and discussion of results, while Section 5 gives the conclusions.

\section{Thruster physical and numerical model}

\subsection{Global source model}
The MEPT considered in this work is a cathode-less RF plasma thruster. A 0D Global Source Model (GSM), as presented in \cite{b:souhairPP,b:AA2022}, was used to evaluate the plasma properties at the exit section of the source tube. 
This volume averaged model is based on the following main assumptions:
(i) cylindrical geometry of the plasma source, (ii) axysimmetric magnetostatic field, and (iii) the magnetic cusps in the source are taken into account by means of semi-empirical corrections.
The governing equations that describe the plasma dynamics are the species density flux and electron power balance
\begin{equation}\label{eqn:Gm_mass}
    \frac{dn_I}{dt}=R_{chem}^I-R_{wall}^I-R_{*}^I+R_{in}^I
\end{equation}
\begin{equation}
    \frac{d}{dt}\left(\frac{3}{2}n_e\langle T_e\rangle\right)=P_{a}'''-P_{chem}'''-P_{wall}'''-P_{*}'''
    \label{eqn:Gm_power}
\end{equation}
where $n_I$ is the number density species $I$ (the subscript $e$ refers to electrons). 
Among these, ions, ground-state neutrals and excited states are present. \textcolor{black}{The excited states of Xe are included in a lumped form \cite{b:souhairPP}, hereafter, the electronic states 5p$^5$6s$^1$ and 5p$^5$6p$^1$ will be referred to as 1S and 2P, respectively, according to Paschen's notation \cite{majorana2021}.}

$R_{chem}$, $R_{wall}$, $R_{*}$, $R_{in}$ are the particle density fluxes related to chemical reactions, wall losses, particle outflow and particle inflow respectively.
Similarly, $P_{a}'''$, $P_{chem}'''$, $P_{wall}'''$ and $P_{*}'''$ are the power densities of the RF power absorbed into the plasma, the power exchanged/lost via chemical reactions, the energy flux losses at the walls, and the power associated to the particle outflow respectively. $\langle T_e \rangle$ is the global-averaged electron temperature. Non-uniformity in the plasma profiles within the source are taken into account by means of semi-empirical coefficients $h_L$ and $h_R$, explained in \cite{b:guaita2022,b:marmuse}.
A more detailed description of the model can be found in \ref{sec:appendixA}.

Given the absorbed power density $P_{a}'''$, propellant mass flow rate $\dot{m}$, and magnetic field $\mathbf{B}$, the GSM solves the system of equations to provide to the PIC model the ion and electron mass flow rate and the electron temperature, respectively $\dot{m}_{i*}$, $\dot{m}_{g*}$ and $T_{e*}$; the subscript $*$ denotes the reference source properties at the exit for the PIC model. 

\subsection{Particle-in-Cell}
\begin{figure*}[!h]
\includegraphics[width=\linewidth]{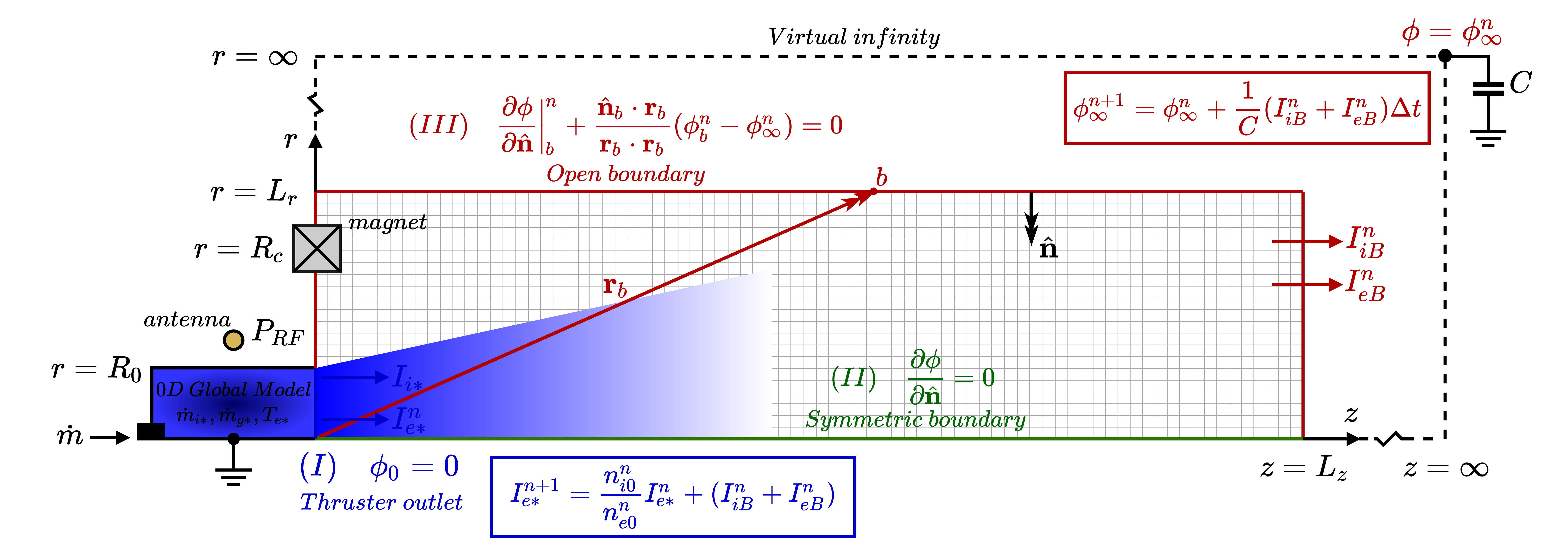}
\caption{\label{fig:domain}Simulation domain and boundary conditions: (I) is the source tube outlet; (II) is the symmetric boundary; (III) is the open boundary.}
\end{figure*}
The plasma topology in the MN has been simulated with an in-house adaption of the fully kinetic axisymmetric 2D3V PIC code \emph{Starfish} \cite{b:andrews2022,b:brieda2012}, which has been experimentally validated in references \cite{andrews2023,b:andrews2022,b:difede2021}. The simulation domain is shown in Fig.~\ref{fig:domain}; it consists of a cylindrical 2D space $(z,r)$. The aim of this work is to simulate purely the plume expansion, hence the source is not included in the domain. As mentioned above, the plasma injection parameters are provided by the GSM. The flow is injected at the source tube outlet ($I$).The external boundaries ($III$) are treated as open to vacuum, connected to the outlet ($I$) via a virtual free-space capacitance which ensures equal ion and electron current streams to the infinity at steady-state. Boundary ($II$) is the axis of symmetry.
Hereafter, the subscripts $*$, $0$, $b$ and $\infty$ shall refer to properties within the plasma source (reference), at the thruster outlet boundary ($I$), at the open boundaries ($III$) and at the virtual infinity respectively. The subscript B shall refer to the integral sum of local properties along the open boundaries. Likewise, the superscripts $+$ and $-$ shall refer to the forward and backward-marching components of the plasma properties.

Electrons  $e$, ions $i$ and neutrals $g$, are all tracked as macro-particles and their motion is propagated with the standard leap-frog Boris algorithm
\begin{equation}
    \frac{\mathbf{v}^{n+1/2}-\mathbf{v}^{n-1/2}}{\Delta t}=\frac{q_I}{m_I}\left( \mathbf{E}^n+\frac{\mathbf{v}^{n+1/2}+\mathbf{v}^{n-1/2}}{2}\times \mathbf{B} \right)
\end{equation}
\begin{equation}
    \frac{\mathbf{r}^{n+1}-\mathbf{r}}{\Delta t} = \mathbf{v}^{n+1/2}
\end{equation}
where $\mathbf{r}^n$ is the particle position at time-step $n$ with velocity $\mathbf{v}^n$, $\Delta t$ is the time-step, $\mathbf{E}$ is the electric field, and $\mathbf{B}$ is the static magnetic field. The currents induced in the plume are negligible \cite{b:jimenez}, hence the magnetic fields can be assumed to be constant. As a result, the electric field $\mathbf{E}$ is curl-free and its potential field $\phi$ can be solved according to the Poisson’s equation
\begin{equation}
    \varepsilon_0\nabla^2\phi = -\rho
\end{equation}
where $\rho = e(n_i - n_e)$ is the local charge density and $\varepsilon_0$ is the vacuum permittivity. At each time-step, the new electric field $\mathbf{E}$ is updated by evaluating
\begin{equation}
    \mathbf{E} = -\nabla\phi
\end{equation}

\subsubsection{Boundary Conditions} 
A reference potential $\phi = 0$ is assigned to the source tube outlet ($I$), the $r = 0$ axis ($II$) is symmetric, hence, the zero-Neumann boundary condition $\partial\phi/\partial\hat{\mathbf{n}}=\mathbf{0}$ is applied there, with $\hat{\mathbf{n}}$ the unit vector normal to the surface. The remaining open boundaries ($III$) are closed by a non-stationary Robin condition introduced by Andrews et al. \cite{b:andrews2022}
\begin{equation}
    \frac{\partial\phi}{\partial\hat{\mathbf{n}}}\bigg|^n_b+\frac{\hat{\mathbf{n}}_b\cdot{\mathbf{r}_b}}{\mathbf{r}_b\cdot\mathbf{r}_b}\left(\phi^n_b-\phi^n_\infty\right) = 0
    \label{eqn:robin}
\end{equation}
where $\mathbf{r}_b$ is the vector distance from the centre of the source outlet (I) to the location on the open boundary (III) and $\phi_\infty$ denotes the free-space plasma potential at infinity. 

At each time-step, particles are injected from the source outlet boundary (I). Since a 0D GSM has been used to estimate the plasma properties, a uniform density is imposed, regarding the velocity, for all species, a Maxwellian distribution is assumed \cite{chen1999}
\begin{equation}
f_{I}^{+}(\mathbf{v}_{I}) = \sqrt{\frac{m_{I}}{2\pi k_B T_{I*}}}\exp \left(-\frac{m_{I}}{2k_B T_{I*}}|\mathbf{v}_{I}-\mathbf{u}_{I}|^2 \right)H(v_{Iz})
\label{eqn:f}
\end{equation}
where $T_{I*}$ is the reference species temperature and $k_B$ is the Boltzmann constant. The drift velocity $\mathbf{u}_I$, imposed along the $z$ direction, is equal to the Bohm speed for both ions and electrons $\mathbf{u}_{i,e} = \langle c_{*},0,0\rangle$, where $c_{*}=\sqrt{k_B T_{e*}/m_i}$. Neutrals possess free-molecular thermal velocity $\mathbf{u}_{g} = \langle \bar{v}_g/4,0,0\rangle$, where $\bar{v}_g=\sqrt{8k_BT_{g*}/\pi m_g}$. $H()$ is the Heaviside function, since only forward-marching distributions ($v_{Iz}>0$) can be imposed.

Ions and electrons (and neutrals) returning to the source are absorbed; ions and neutrals reaching the open boundaries are also absorbed. For electrons at the open boundaries, an energy-based reflection criterion is used to account for the trapped population returned by the ambipolar potential drop \cite{b:andrews2022}. The kinetic energy of electrons $KE_{e}$ is compared to the trapping potential $PE_b$ on the boundary node.
\begin{equation}
    KE_{e} = \frac{1}{2}m_e|\mathbf{v}_e|^2
\end{equation}
\begin{equation}
    PE_{b}=e(\phi_b-\phi_\infty)
\end{equation}
If $KE_{e}<PE_{b}$ the electron is trapped, so it is reflected back with velocity $-\mathbf{v}_{e}$. Else, it is a free electron to be removed from the domain.

From this energy-based criterion, there is a value of $\phi_\infty$ that reflects sufficient electrons to maintain a current-free plume. Therefore, the value of $\phi_\infty$ is self-consistently controlled via a virtual free-space capacitance $C$
\begin{equation}
    \phi_\infty^{n+1}=\phi_\infty^n+\frac{1}{C}(I_{iB}^n+I_{eB}^n)\Delta t,
    \label{eqn:control}
\end{equation}
where $I_{iB}$ and $I_{eB}$ are the global sum ion and electron currents leaving the open boundaries ($III$). This method inherently guarantees that, once at steady-state, the ion and electron currents streaming to infinity are equal.

During the transient, any non-zero net current leaving the open boundaries ($III$) must be re-injected into the domain via the thruster outlet ($I$) to complete the circuit. In addition, the injected electron current $I_{e*}$ is controlled in order to enforce the quasi-neutrality condition at the source outlet ($I$). From these considerations, the following conditions are imposed to particles injected at boundary ($I$). Ions are injected with a constant current given by $I_{i*} = en_{i*}c_{*}A_0$, where $A_0$ is the area of the source outlet. The injected electron current is updated each time step according to 
\begin{equation}
{I_{e*}^{n+1}} = (I_{iB}^n+I_{eB}^n) + \frac{n_{i0}^n}{n_{e0}^n}{I_{e*}^n}
\label{eqn:QN}
\end{equation}
where the first term completes the circuit and the second enforces the quasi-neutrality. Considering that injected electrons are Maxwellian, the initial value of the current is set to be ${I_{e*}^0}=-en_{*}(\bar{v}_{e*}/4+c_{*})A_0$. The neutral flux is imposed as $\Gamma_{g*}=n_{g*}\bar{v}_g/4A_0$. 
Since ions are only accelerated outward by the electric field, and neutrals diffuse guided by the outer pressure gradient, no treatment is required for a possible backflow of heavy particles.

For a more detailed description of this PIC formulation, including validation, derivation of the Robin condition, domain-independence studies, and a numerical sensitivity analysis, the reader is directed to the previous work of Andrews et al. \cite{b:andrews2022,b:Numerical}.

\subsubsection{Collisions}

The model considers electron-neutral elastic and inelastic (i.e., ionisation and excitation) scattering, ion-neutral scattering, ion-neutral charge exchange (CEX), Coulomb collisions and an equivalent anomalous collisionality. Neutrals are at ground state (i.e. not excited) by default and any excited states are assumed to decay immediately. Collisions between different species are handled with the Monte Carlo Collisions (MCC) method \cite{MCC}, while same species interactions were considered with full Direct Simulation Monte Carlo (DSMC) treatment \cite{bird2013dsmc}. The list of collisions implemented in the numerical suite and their respective references are provided in Table~\ref{Tab:Cross-sections}. Anomalous collisions are taken into account through the empirical Bohm model \cite{b:Bohm}, with a diffusion coefficient $\alpha_{an} = 1/64$. 
\begin{table}[!b]
\centering
\begin{tabular}{ccc}
\hline
\textbf{Reactions} & \multicolumn{2}{c}{\textbf{Reaction-type}}\\
\hline
$e + e \longrightarrow e + e$         & Coulomb scattering & \cite{weng1990method} \\
$e + Xe^+ \longrightarrow e + Xe^+$      & Coulomb scattering & \cite{b:szabo} \\
$e + Xe^+  \longrightarrow e + Xe^+$      & Bohm collision     & \cite{b:esipchuck1976}\\
$e + Xe   \longrightarrow 2e + Xe\textcolor{black}{^+}$       & Ionization         & \cite{b:szabo} \\
$Xe^+ + Xe \longrightarrow Xe^+ +Xe$      & Momentum exchange  & \cite{b:szabo} \\ 
$Xe^+ + Xe \longrightarrow Xe + Xe^+$     & Charge exchange    & \cite{b:szabo} \\
$e + Xe   \longrightarrow e + Xe$        & Elastic scattering & \cite{b:szabo} \\
$Xe + Xe   \longrightarrow Xe + Xe$       & Elastic scattering & \cite{b:szabo} \\
$e + Xe  \longrightarrow e + Xe_{1S}$  & Excitation &  \cite{b:souhairPP}\\
$e + Xe   \longrightarrow e + Xe_{2P}$   & Excitation &  \cite{b:souhairPP}\\
\hline
\end{tabular}
\caption{\label{Tab:Cross-sections}Interactions cross-sections.}
\end{table}
When considering electron-neutral excitation interactions, taking into account the dynamics of each fine-structure excitation state would represent an unbearable computational effort. Instead, a lumping of the energy levels is performed, implemented according to Souhair et al. \cite{b:souhairPP}. 
\textcolor{black}{Excited states are lumped to divide resonant and metastable species as well as 1S and 2P. This lumping methodology requires the assumption of local thermodynamic equilibrium (LTE) according to McWhirter \cite{b:mcWhirter}. For the temperatures handled in this work ($T_e< 20$~eV), and the maximum energy jump, the formula gives a threshold value of $n_e > 10^{16}$ m$^{-3}$ \cite{b:souhairPP}. Since more than 95\% of the excitation collision takes place where such density criterion is satisfied, LTE is a reasonable assumption within the scope of this paper.}

If a collision takes place, the electron is assumed to lose energy equal to the lumped excitation energy ($\Delta U_{IJ}=U_J-U_I$), i.e. $E_e^{n+1}=E_e^n-\Delta U_{IJ}$, and its speed is reduced accordingly. \\
The lumped excitation cross-section can be evaluated by means of an adapted version of the procedure defined in \cite{b:souhairPP}:
\begin{equation}
    \sigma_{ex_{IJ}} = \sum^{N_i}_i \left[ \frac{\sum^{N_j}_j \sigma_{ex_{ij}}}{\sum^{N_i}_k\frac{g_k}{g_i}\exp\left(-\frac{U_k-U_i}{k_B T_g}\right)} \right]
\end{equation}
where $\sigma_{ex_{IJ}}$ represents the cross-section for the lumped state $I$ to $J$,  $U_k - U_i$ is the energy difference between the k-th and the i-th fine-structure levels, $g_k$ and $g_i$ are the statistical weights, and $T_g$ is the neutral gas temperature.\\

\subsubsection{Numerical acceleration}
The simulation time-step should be small enough to resolve plasma frequency $\omega_{pe}=\sqrt{n_ee^2/\varepsilon_0m_e}$; this, combined with fine mesh requirements where spacing is imposed by the Debye length $\lambda_D=\sqrt{\varepsilon_0k_BT_e/n_ee^2}$, results in an unmanageable computational load.\\
In order to reduce this effort, two different numerical acceleration approaches have been adopted \cite{b:SzaboNumerical}. The vacuum permittivity is increased by a factor $\gamma^2$; the Debye length then increases by $\gamma$, allowing for a less refined mesh and fewer macro-particles. Plasma frequency also slows by a factor $1/\gamma$, hence a larger time-step may be chosen.\\
Second, the heavy particle mass (i.e. $m_i$ and $m_g$) is reduced by a factor $f$; this increases their velocity by $\sqrt{f}$.
The combination of these reduces the simulation time by approximately $\gamma^2\sqrt{f}$. Reference \cite{b:andrews2022} reports a sensitivity analysis on the values of $\gamma$ and $f$ used in the PIC model. All the results that are shown in this work are presented unscaled. 

\section{Performance indicators}

In order to define the propulsive performance and to analyse the power balance of the MN, maps of macroscopic quantities (e.g., density, velocity, temperature) are computed by integrating the moments of the distribution functions computed with the PIC \cite{b:chen1984}. In the following, the relations used to derive performance indicators (e.g., thrust) from these maps are presented.

\subsection{Thrust}

The total thrust $F$ has been evaluated as the combination of the force produced by the plasma expansion through the source tube outlet ${F}_0$ and the force imparted by the MN ${F}_{j\times B}$,
\begin{equation}\label{eq:thrust}
    {F}={F}_0 + \underbrace{\iiint\limits_V -j_{e\theta} B_r\ dV}_{F_{j \times B}}
\end{equation}
where $j_{e\theta} = -en_eu_{e_\theta}$ is the azimuthal electron current density ($u_{e_\theta}$ being the azimuthal electron speed), $B_r$ is the radial component of the magnetic field $\mathbf{B}$, and $V$ is the PIC domain volume. \\
The plasma source contribution to thrust ${F}_0$ can be obtained by computing Eq.~\ref{eq:mflux} at the exit section of the plasma source $dS=dA_0$ \cite{b:Ahedo2013}.
\begin{equation}\label{eq:mflux}
    {F} =\oiint\limits_{S} \sum_I \left( m_I n_I \mathbf{u}_I\mathbf{u}_I + p_I\mathbf{\overline{\overline I}} +{\overline{\overline\pi}}_I\right)\cdot \hat{\mathbf{n}} ~dS
\end{equation}
$p_I=k_Bn_IT_I$ is the scalar pressure, $\mathbf{\overline{\overline I}}$ is the identity tensor, and ${\overline{\overline\pi}}_I$ is the stress tensor \cite{b:bittencourt2004}.

\subsection{Power balance}

The energy equation \cite{b:bittencourt2004}, for a steady state flow reads
\begin{equation}\label{eq:balance}
    P_{kin,I} + P_{T,I} + Q_I + P_{coll} = 0
\end{equation}
the different terms represents the species kinetic ($P_{kin,I}$) and thermal convection powers ($P_{T,I}$), the heat flux ($Q_I$) and the overall power losses due to collision between particle $I$ and a particles of a different species.
The following power contributions have been considered in order to evaluate the propulsive efficiencies:
\begin{align}
    P_{kin,I}  &= \oiint\limits_S\frac{1}{2}n_Im_Iu_I^2\mathbf{u}_I\cdot\mathbf{\hat{n}}~dS\\
    P_{T,I}  &= \oiint\limits_S\frac{5}{2}n_IT_I\mathbf{u}_I\cdot\mathbf{\hat{n}}~dS
\end{align}
$P_*$ is the source tube exhaust power as provided by Eq.~\ref{eqn:Gm_power} of the GSM; it can also be given by,
\begin{equation}
    P_* \approx \sum_I P_{kin,I}(dA_0) + \sum_I P_{T,I}(dA_0) + \sum_I Q_k(dA_0)
\end{equation}
For collisional interactions, Eq.~\ref{eq:losses} allows the computation of the power involved in the particular elastic or inelastic interaction
\begin{equation}\label{eq:losses}
    P_{coll_{IJ}} = \iiint\limits_V n_I\mathbf{\nu}_{IJ}\Delta E_{IJ}~dV
\end{equation}
for interaction between species $I$ and $J$. $\nu$ is the collision frequency and $\Delta E_{IJ}$ is the energy variation that depends on the particular reaction.
It is important to note that, considering equation Eq. \ref{eq:balance}, for a particular species, both elastic and inelastic interactions have the effect of reducing (or increasing) the power associated to its flow, as it is exchanged to a different species. This does not necessarily mean that the overall power of the expanding plasma decreases. When considering the working principle of a magnetic nozzle, it appears evident how no direct interaction with neutrals is present, therefore, little to no contribution to thrust is provided by the cold gas. As a consequence, all the power that is exchanged towards neutrals (belonging to the plume or the background) will, from now on, be considered (and referred to) as a loss, also done in \cite{wachs2020background,emoto2021}. \textcolor{black}{Regarding the charge-exchange interaction between ions and background neutrals, whenever a fast ion collides with a neutral particle, the result will consist in the generation of a fast atom and a slow ion. Since no thrust contribution is provided by neutrals outside of the source, given their inability to interact with the thruster via the MN, the net momentum gain of the $n_g$ population does not translate in a performance gain. Adversely, having a fast ion instantly reducing its velocity causes the charged particle momentum flux to decrease, negatively affecting the net thrust.}

For the generic elastic collision, $\Delta E_{IJ}$ coincides with the kinetic energy of the reduced system:
\begin{equation}\label{eq:elasticLoss}
\Delta E_{IJ}^\mathrm{el} =\frac{1}{2} m_r (|\mathbf{u}_I| - |\mathbf{u}_J|)^2
\end{equation}
where $m_r$ is the reduced mass. In the case of inelastic collision losses, the energy term is instead fixed according to the relevant threshold energy (e.g. ionisation and excitation).

The following three main efficiencies can then be defined for the MN:
\begin{align}
    &\eta_{conv}^{} = \frac{P_{kin,i}}{P_*}\\
    &\eta_{div}^{}  = \frac{P_{kin,i}^{(z)}}{P_{kin,i}}\\
    &\eta_{MN}^{}   = \frac{\sum_I \left(P_{kin,I}^{(z)} + P_{T,I}\right)}{P_*}
\end{align}
where $\eta_{conv}^{}$ is the internal to ion-kinetic energy conversion efficiency, $\eta_{div}^{}$ is the ion divergence efficiency (i.e.~a measure of ion confinement), and $\eta_{MN}^{}$ represents the overall MN efficiency \cite{b:Ahedo2013}. 

\section{Results and discussion}

\subsection{Thruster specifications}

The thruster considered in this work is a laboratory prototype, derived from REGULUS-150 \cite{andrews2023}.
The plasma source is a cylindrical tube of length $L_s = 0.100$~m and radius $R_s = 0.0085$~m. The on-axis magnetic field intensity at the MN throat is $B_0 \approx 450$~G; the field topology is given in Fig.~\ref{fig:neutrals}~(a). The RF antenna is a 0.02~m long five-turn copper coil, with a wire width of $w_A = 0.002$~m \cite{b:AA2022}. Xenon propellant is delivered to the injector with mass flow rate $\dot{m}=0.25$~mg/s. The antenna coupling efficiency is assumed to be $\eta_{RF}\approx 0.7$.

\subsection{Operating condition and neutral gas density}

\begin{figure}[!b]
    \centering
    \includegraphics[width=0.7\linewidth]{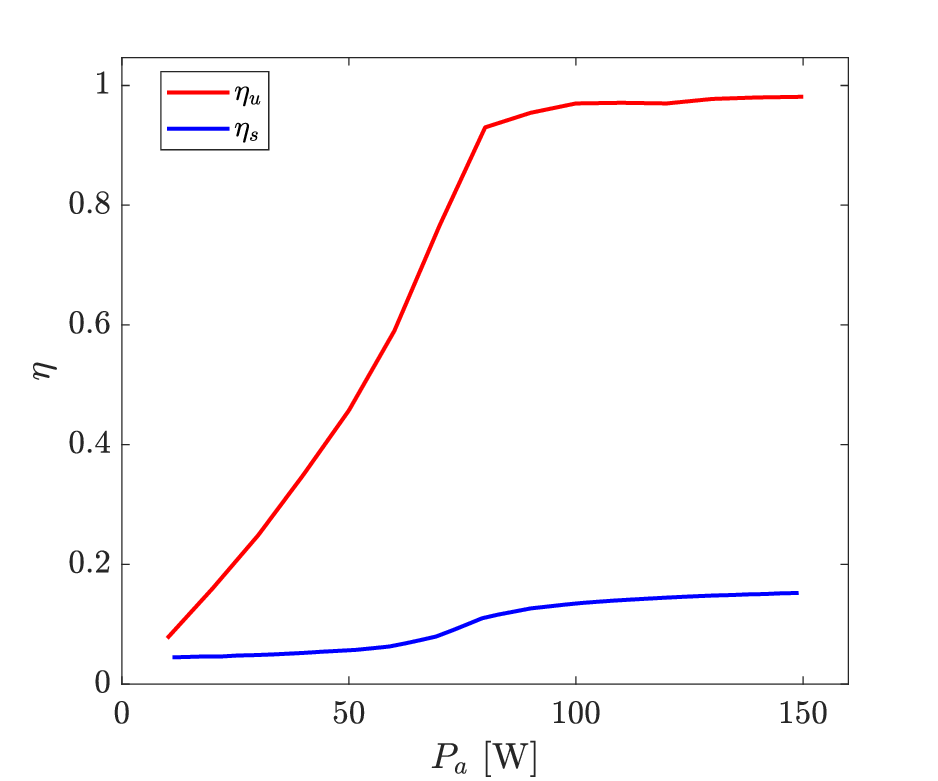}
    \caption{Mass utilisation efficiency $\eta_u$ and source efficiency $\eta_s$ as function of the absorbed power $P_a$.}
    \label{fig:muEff}
\end{figure}

The MN has been simulated at a range of absorbed powers, with source reference properties from the GSM given in Table~\ref{Tab:Inputs}. 
\begin{table}[!hb]
\centering
\begin{tabular}{cccccc}
\hline
$P_a$ [W] & T$_{e*}$ [eV] & n$_{i*}$ $\times10^{18}$ [m$^{-3}$] &  n$_{g*}$ $\times10^{19}$ [m$^{-3}$] & $\eta_u$ & $\eta_s$\\ 
\hline 
60   & 4.20  & 1.22 & 2.75 & 0.59 & 0.059\\
90   & 8.32  & 1.41 & 0.30 & 0.95 & 0.126\\
130  & 13.76 & 1.12 & 0.15 & 0.98 & 0.148\\
150  & 16.34 & 1.03 & 0.13 & 0.98 & 0.153\\
\hline
\end{tabular}
\caption{\label{Tab:Inputs}Plasma parameters at the source tube exit for different absorbed power.}
\end{table}
For the main discussion hereafter, the $P_a=60$~W and $P_a=130$~W cases are focused upon: the former is a low-temperature ($T_{e*}<\Delta E_{ion}$=12.1 eV), high neutral density ($n_{g*}>n_{i*}$) mode; the latter is a higher temperature ($T_{e*}>\Delta E_{ion}$), low neutral density ($n_{g*}<n_{i*}$) regime. This pronounced variation in the neutral concentration can be explained by looking at Fig.~\ref{fig:muEff}. Defining the mass utilisation efficiency ($\eta_u = \dot{m_{i*}}/\dot{m}) $ of the RF-source as the ratio between the ion mass flow rate and the overall flow rate, it is quite evident how below the threshold of 80 W, the ion-to-neutral ratio quickly drops, resulting in a massive difference in the outlet parameters. Fig.~\ref{fig:muEff} also shows the source efficiency $\eta_s = P_*/P_a$ defined as the ratio between the power at the exhaust of the source chamber and the power absorbed by the plasma.

Background pressure is included in the PIC by setting a fixed uniform neutral distribution $n_{g}^\mathrm{back}$ via the ideal gas law
\begin{equation}
    p_\mathrm{back} = n_{g}^\mathrm{back}k_BT_{g}^\mathrm{back}
\end{equation}
at a nominal temperature of $T_{g}^\mathrm{back}=300$~K. Starting from the vacuum scenario $n_{g}^\mathrm{back} = 0$~m$^{-3}$ at $p_\mathrm{back} = 0$~Pa, this leads to $n_{g}^\mathrm{back} = 2.51\times10^{18}$~m$^{-3}$ at $p_\mathrm{back} = 10^{-2}$~Pa.

Before advancing in the study, a quick consideration about the hypothesis of uniform background distribution shall be made. When dealing with vacuum chamber gas distribution the gradient \textcolor{black}{of the neutral gas pressure} is generally dependent on the individual chamber arrangement. Moreover, several experimental measures \cite{boyd2003,cai2006} and DSMC \cite{passaro2004} simulations have proven the neutral map to be basically constant around the plume expansion region, as long as the vacuum chamber is well dimensioned for the experiment. Therefore, the assumption of a constant background allows for relatively valid results while not losing the generality of the study.
\begin{figure}[!t]
\centering
    \centerline{\includegraphics[width=1.2\linewidth]{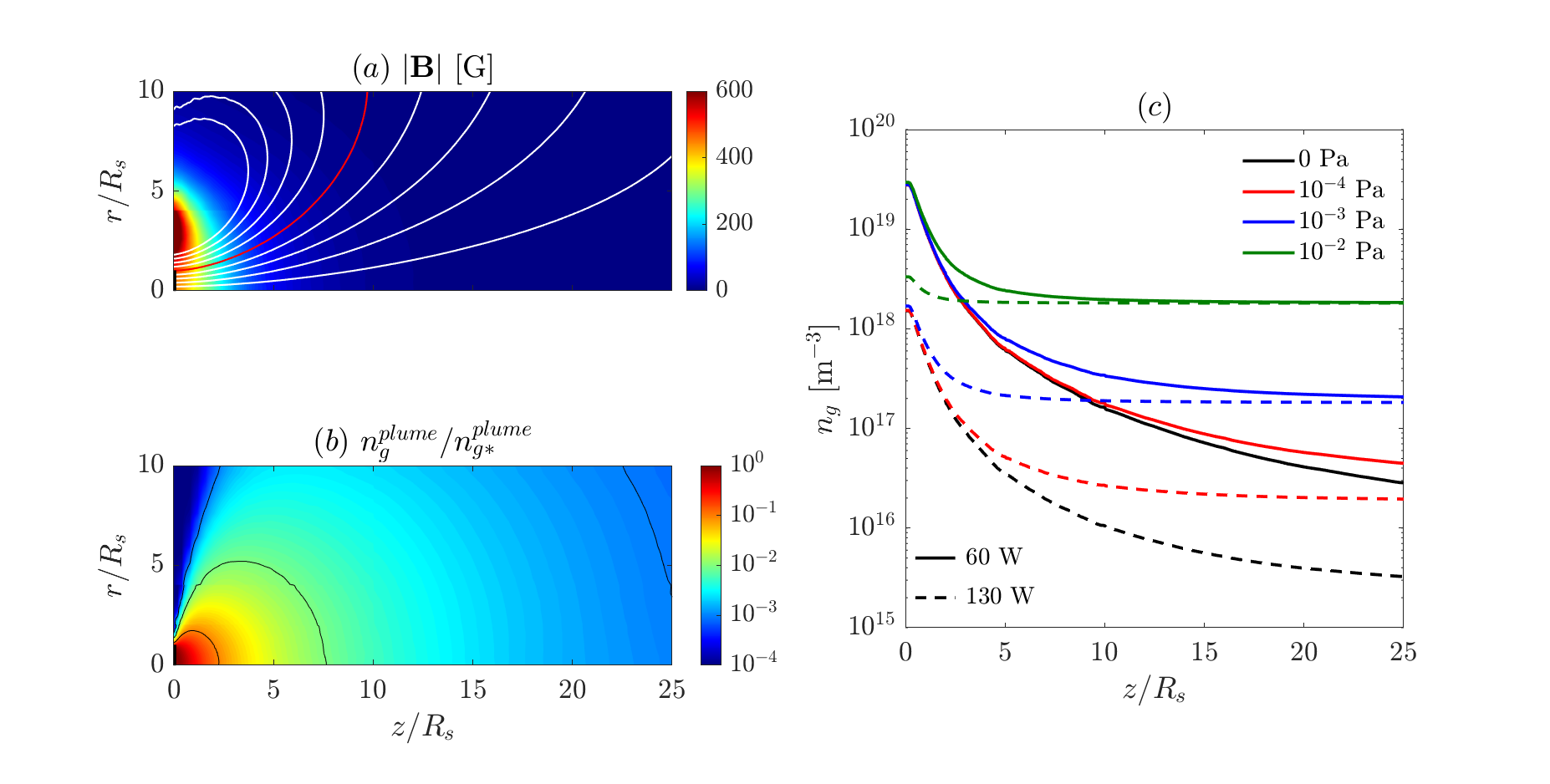}}
    \caption{(a) Magnitude of the magnetic flux density $|\mathbf{B}|$. The outermost magnetic field line connected to the source is given by the red contour; (b) Normalised neutral gas density from the plume discharge $n_{g}^\mathrm{plume}/n_{g*}^\mathrm{plume}$; (c) On-axis total neutral gas density $n_{g}^\mathrm{tot}$ for $P_a=$ 60 W and 130 W.}
    \label{fig:neutrals}
\end{figure}
Fig.~\ref{fig:neutrals} (b) shows the normalised neutral density from the thruster plume $n_{g}^\mathrm{plume}$, pre-computed with the DSMC method. The on-axis combination of both plume and background density $n_{g}^\mathrm{tot}=n_{g}^\mathrm{plume}+n_{g}^\mathrm{back}$ is then given in Fig.~\ref{fig:neutrals} (c) for $P_a=60$ and 130 W for $p_\mathrm{back}=0-10^{-2}$ Pa. Due to the lower $n_{g}^\mathrm{plume}$ of the high power case (from greater ionisation efficiency) there is a more decisive variation of the total density with changing facility pressure.

\subsection{Plasma profiles}

\begin{figure*}[!t]
    \centering
    \includegraphics[width=1\linewidth]{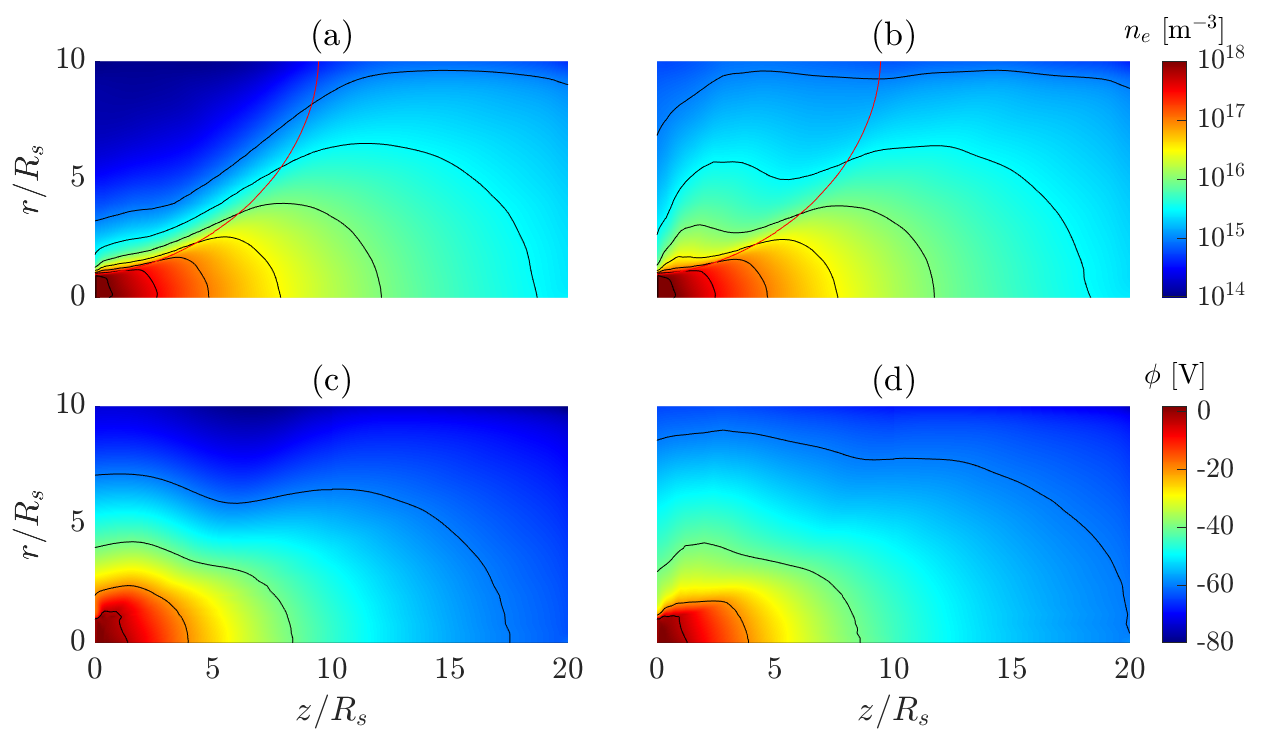}
    \caption{2D profiles of (a)(b) electron number density $n_e$ (c)(d) plasma potential $\phi$ for 130 W for background pressure 0 Pa (left) and 10$^{-2}$ Pa (right). The red line indicates the  outermost magnetic field line connected to the plasma source.}
    \label{fig:2Dfields}
\end{figure*}
In order to gain insight into the plume topology, the 2D plasma profiles are provided in Fig.~\ref{fig:2Dfields} for the 130 W case. Figs.~\ref{fig:2Dfields} (a) and \ref{fig:2Dfields} (b) show electron number density distribution for vacuum conditions and $10^{-2}$ Pa respectively. At low-pressure, the plume is highly-collimated and the expansion is confined within the outermost magnetic field line (OMFL) connected to the source. Only a small fraction of electrons is able to escape the confinement of the MN; this is due to their high energy. In Fig.~\ref{fig:2Dfields} (b), a more diffusive behaviour appears as the electrons that manage to escape from the OMFL are now more than just the high energy tail of the discharge Maxwellian distribution.

Despite this enhanced cross-field diffusion, the density does not considerably reduce in the downstream region of the plume. This is because the increased neutral background produces significant in-plume ionisation. Low-energy ions are formed in the near-field and are therefore heavily influenced by the radial electric field; the ions accelerate radially and form a secondary density cloud similar to that seen from CEX. With the possible exception of ions that have experienced CEX collisions, ions born in the near-field plume region also experience a smaller potential drop than ions born in the source tube. The mean ion speed within the fluid may be significantly decreased by this newly ionised population, serving as an effective drag term on the ions \cite{b:andrews2022}.

Figs.~\ref{fig:2Dfields} (c) and (d) give the plasma potential. Note that the enhanced radial potential peak observed in \cite{b:andrews2022} does not appear in Fig.~\ref{fig:2Dfields} (c) since the near-exit neutral density at 130 W is insufficient to produce the necessary ion-neutral collisions, but a mild ion-confining potential barrier is still present. This structure completely fades at $10^{-2}$ Pa in Fig.~\ref{fig:2Dfields} (d). In particular, the plasma potential appears to have a smoother gradient in all the directions. The downstream potential drop is then reduced by approximately 5 V by the domain boundaries.  

\begin{figure}[!t]
    \centering
    \includegraphics[width=1\linewidth]{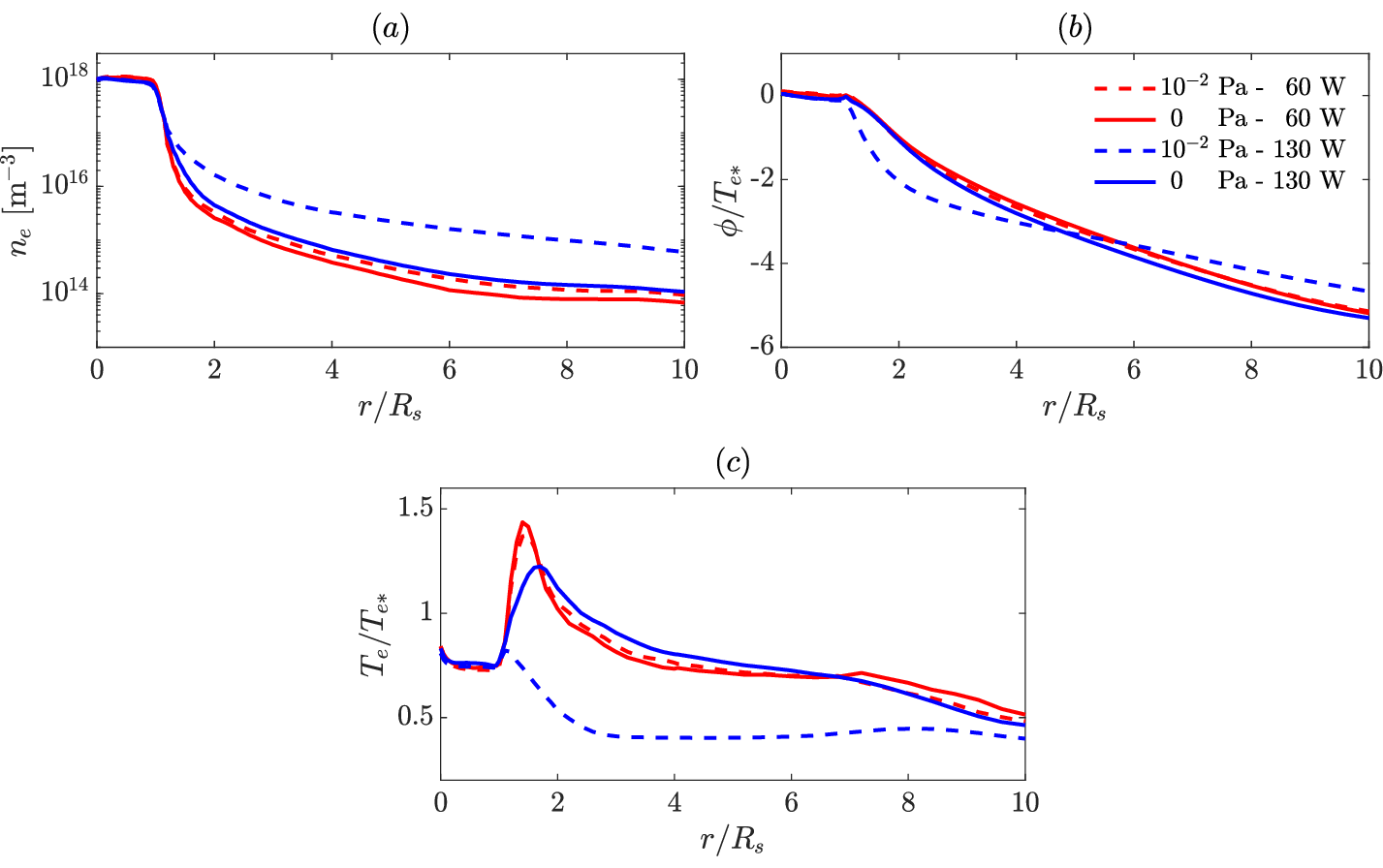}
    \caption{Comparison between (a) radial ion number density (b) radial plasma potential (c) electron temperature at different background pressure for operating power 60 W and 130 W.}
    \label{fig:radialProfiles}
\end{figure}
Fig.~\ref{fig:radialProfiles} provides the radial profiles of $n_e$ and $\phi$ at $z = 5$~mm. Fig.~\ref{fig:radialProfiles} (a) shows the radial electron density; it is quite evident how collisions affect the diffusion more in the high-pressure scenario, where the number density variation at the radial periphery increases by a factor of 10. The low-power case seems to be almost unaffected; this can be explained by looking once again at Fig.~\ref{fig:neutrals}. The radial profiles have been chosen at the axial position of $z = 5$~mm. At this distance from the source tube exit, for low-power, the neutral gas density from the source is greater with respect to the background (n$_{g*}^\mathrm{plume}\sim$ 10$^{19}$ m$^{-3}$), therefore the effect of additional neutrals is negligible. The same cannot be stated for the high-power case; in fact, with the neutral density of the same magnitude as the ambient one, the additional scattering caused by interactions highly affects the plume topology.\\
Considering Fig.~\ref{fig:radialProfiles} (b) it is quite evident how the radial component of $\phi$ presents a different behaviour. The potential drop in the high-pressure high-power case is far more accentuated, hence the ion confinement is less effective \cite{b:andrews2022}. As a consequence, a more pronounced secondary expansion is present and a less-collimated plume can be observed. The tendency of the plasma beam to spread in the radial direction (i.e. the presence of a less effective magnetic confinement) can be seen also in Fig.~\ref{fig:radialProfiles} (c), where the electron temperature is shown as a function of the radial position. It is quite evident of the population of electrons that is able to escape the MN confinement is an high temperature one. The behaviour of this high-energy escaping population is barely affected by the pressure variation when the operating power is limited (the overall electron number density is almost constant along the radius), whereas, at increasing power, the magnetic confinement is overcome also by low-energy particles, that are able to pass the OMFL due to the radial momentum component induced by collision. As a consequence, the temperature peak becomes extremely limited and the electron population quickly cools down until a plateau is reached.
\begin{table}
\centering
\begin{tabular}{ccc}
\hline
& \multicolumn{2}{c}{$\phi_{\infty}/T_{e*}$}\\ 
\hline
$p_\mathrm{back}$ [Pa] &  $P_a$ = 60~W & $P_a$ = 130~W\\ 
\hline
$0$       & -7.0 & -6.9\\
$10^{-4}$ & -7.0 & -6.9\\
$10^{-3}$ & -7.0 & -6.8\\
$10^{-2}$ & -6.8 & -6.3\\
\hline
\end{tabular}
\caption{\label{Tab:PhiInf}Normalised plasma potential at infinity $\phi_{\infty}$ for increasing background pressure at different operation power $P_{a}$.}
\end{table}
The values of plasma potential at infinity can be found in Table~\ref{Tab:PhiInf}. It is clear the potential drop reduces in magnitude with increasing pressure, decreasing the ambipolar ion acceleration, resulting in a lower thrust.
The detriment is larger for the high-power case ($\sim$10\%), starting with $\phi_{\infty}/T_{e*} \simeq -6.9$ up to $\phi_{\infty}/T_{e*} \simeq -6.3$; adversely, the low-power potential drop reduction is essentially negligible ($<$3\%).

\begin{figure}[!ht]
    \centering
    \includegraphics[width=0.7\linewidth]{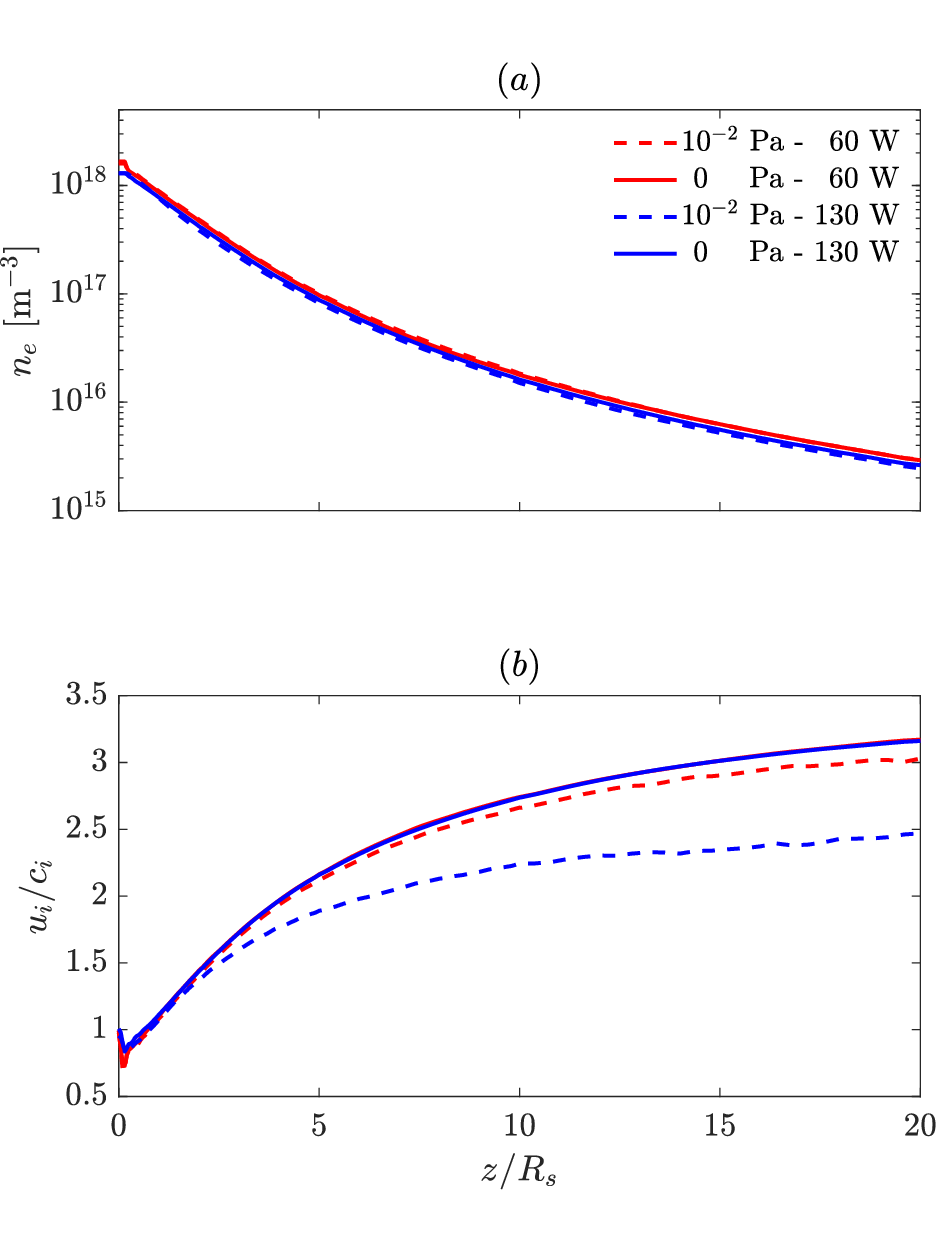}
    \caption{Comparison between (a) axial electron number density and (b) normalised axial ion velocity at different background pressure for absorbed power 60 W (continuous line) and 130 W (dashed line)}
    \label{fig:axialProfiles}
\end{figure}

Fig.~\ref{fig:axialProfiles} shows the axial profiles of $n_e$ and $u_{i}$. Considering Fig.~\ref{fig:axialProfiles} (a) it is fairly reasonable to state that the on-axis electron distribution is not affected by the \textcolor{black}{background neutral population}.
The effect of the decreasing potential drop in the axial direction is quite evident when looking at Fig.~\ref{fig:axialProfiles} (b), where the axial ion velocity $u_i$ is presented normalised by the inlet Bohm speed $c_i = \sqrt{k_b T_{e*}/m_i}$. The initial decrease of the ion velocity is due to collisions (e.g. charge exchange interactions) taking place right after the particles injection in the domain.
As expected, the axial variation of the low-power cases is small ($<$5\%). Instead, the difference becomes quite consistent at 130 W ($\sim$22\%), where the ion velocity begins to diverge from the general trend at $z = 20$ mm.

\subsection{Propulsive Performance}\label{sec:performance}

The analysis of the plasma profiles highlighted how a substantial difference is introduced at high-pressure. In this section, the effect of such variation on the propulsive parameters will be investigated.
\begin{table}
\centering
\begin{tabular}{ccccc}
\hline
& \multicolumn{2}{c}{$P_a$ = 60~W} & \multicolumn{2}{c}{$P_a$ = 130~W}\\ 
\hline
$p_\mathrm{back}$ [Pa] & F/F$_0$ & $\varepsilon$ & F/F$_0$ & $\varepsilon$\\ 
\hline 
0         & 1.44 &  0     & 1.49 &  0    \\
10$^{-4}$ & 1.44 & -0.13  & 1.49 & -0.09 \\
10$^{-3}$ & 1.43 & -0.33  & 1.47 & -1.32 \\
10$^{-2}$ & 1.42 & -1.20  & 1.29 & -13.63\\
\hline
\end{tabular}
\caption{\label{Tab:Thrust}Normalised thrust $F/F_0$ and percentage variation $\varepsilon$ for increasing background pressure at different operation power $P_a$.}
\end{table}

The propulsive performance of the cathode-less RF thruster has been reported in Table~\ref{Tab:Thrust}. It is quite evident how the high-power case is more affected by the increase of the facility pressure, showing a decisive performance degradation of $\varepsilon = 13.63~\%$ at $p_\mathrm{back} = 10^{-2}$~Pa, where $\varepsilon$ is defined as
\begin{equation}
    \varepsilon = \frac{\|{F - F_{p_\mathrm{back} = 0~ \mathrm{P_a} }}\|}{\|{F_{p_\mathrm{back} = 0~ \mathrm{P_a}}}\|}\times 100   
\end{equation}
The 60~W case presents a negligible thrust reduction until the background neutral density becomes closer to the order of the gas emitted by the source (i.e. 10$^{18}$~m$^{-3}$). Even in this case, however, the thrust reduction is more limited with respect to the high-power scenario.\\
\begin{figure}[!ht]
    \centering
    \includegraphics[width=0.7\linewidth]{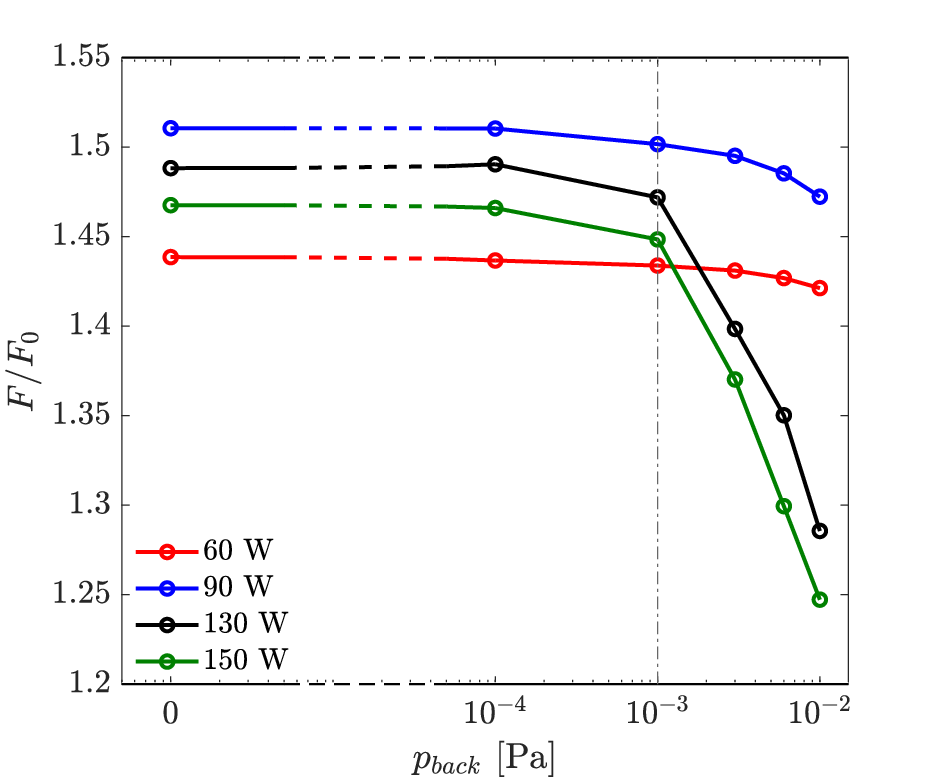}
    \caption{Normalised thrust for increasing background pressure $p_\mathrm{back}$ at different operating powers.}
    \label{fig:thrust}
\end{figure}
\\
The trend of the normalised thrust for the four power levels is shown in Fig.~\ref{fig:thrust}. It is quite evident how the thruster performance is not affected by neutrals in the same way for all the various cases. The value of $10^{-3}$ Pa seems to be a threshold over which the performance drop becomes non-negligible when the thruster is operated at a relatively high power (e.g. 130 W).

\begin{figure}[!t]
    \centering
    \includegraphics[width=0.7\linewidth]{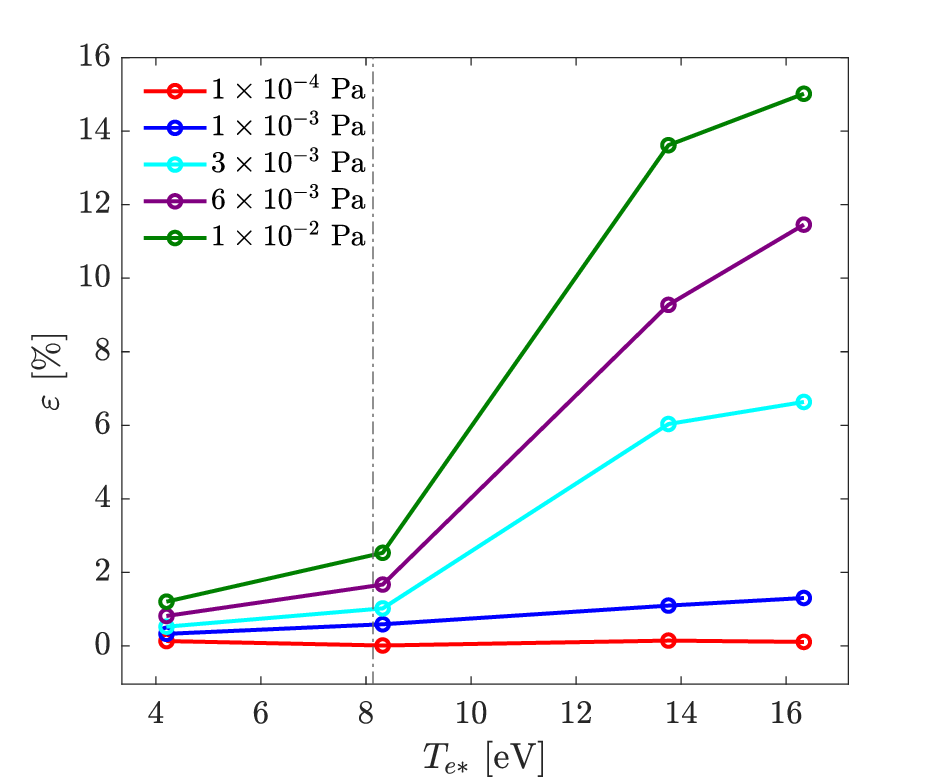}
    \caption{Thrust loss $\varepsilon$ for increasing plasma temperatures at different background pressure levels.}
    \label{fig:thrustLoss}
\end{figure}

Plotting the performance loss $\varepsilon$ as a function of the source plasma temperature $T_{e*}$ for different pressure levels yields Fig.~\ref{fig:thrustLoss}. While low neutral density cases are characterised by a limited $\varepsilon$ throughout the temperature range, a different scenario emerges when the pressure increases. It is quite evident how the slope changes when larger values of $T_e$ are involved. This threshold level could be identified as $T_e \sim 8$~eV (vertical dashed line in Fig.~\ref{fig:thrustLoss}), where excitation and ionisation rates become non-negligible. In fact, as it will be presented in the following section, the ionisation phenomenon becomes of main relevance when considering the higher power cases (i.e. 130 W and 150 W); adversely, at 60 W and 90 W the average energy of electrons is not enough to trigger ionisation interactions. If the electron temperature exceeds the excitation/ionisation energies of the neutral species, facility effects become non-negligible; bearing in mind that the neutral gas plume from the source tube also plays a crucial role in the sensitivity of the thruster to the cold gas background. In general, facility effects must be considered if electron energies in the plume approach levels for significant inelastic collisions.
\\
\begin{table}
\centering
\begin{tabular}{ccccccc}
\hline
 & \multicolumn{3}{c}{$P_a$ = 60~W} & \multicolumn{3}{c}{$P_a$ = 130~W}\\
\hline
$p_\mathrm{back}$ [Pa] & $\eta_{conv}^{}$ &  $\eta_{div}^{}$ & $\eta_{MN}^{}$ & $\eta_{conv}^{}$ &  $\eta_{div}^{}$ & $\eta_{MN}^{}$ \\
\hline
0         & 0.71 & 0.63 & 0.57 & 0.72 & 0.62 & 0.55\\ 
10$^{-4}$ & 0.72 & 0.63 & 0.57 & 0.72 & 0.62 & 0.55\\ 
10$^{-3}$ & 0.71 & 0.63 & 0.56 & 0.71 & 0.60 & 0.54\\ 
10$^{-2}$ & 0.66 & 0.62 & 0.52 & 0.63 & 0.48 & 0.42\\ 
\hline
\end{tabular}
\caption{\label{Tab:Efficiencies}Engine efficiencies for increasing background pressure at different operation power $P_a$.}
\end{table}
Considering now the previously defined efficiencies, Table~\ref{Tab:Efficiencies} shows how they change as the neutral density increases. More in detail, the low-power case thermal conversion and the overall MN efficiencies seems to be slightly affected only at the highest pressure levels, both presenting a decrease of 0.05. In contrast, the ion divergence efficiency looks quite unscathed, with a reduction of 0.01.
As is conceivable, increasing the background pressure at high-power levels gives rise to a harsh drop of the thruster performance parameters, namely 0.09 for $\eta_{conv}^{}$, 0.14 for $\eta_{div}^{}$ and 0.13 for $\eta_{MN}^{}$.\\
\begin{figure}[!h]
    \centering
    \includegraphics[width=0.7\linewidth]{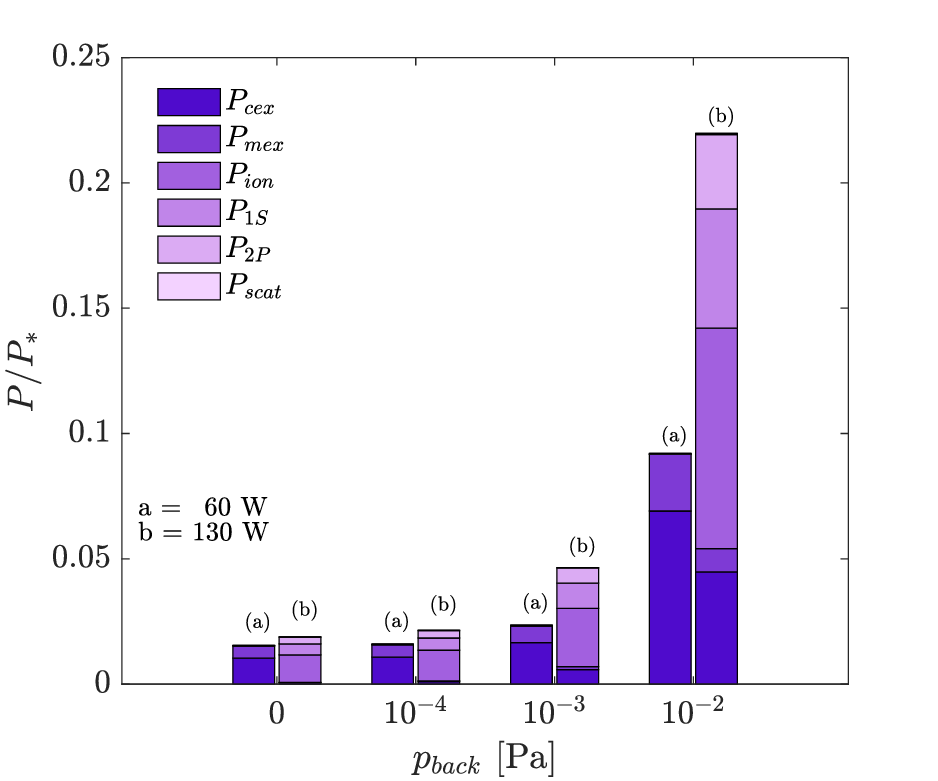}
    \caption{Normalised power losses caused by collision for increasing background pressure $p_\mathrm{back}$ at different operating power (a) 60 W and (b) 130 W.}
    \label{fig:powerBar}
\end{figure}
\\
Fig.~\ref{fig:powerBar} shows the different normalised collisional power losses, i.e. the power that ions and electrons lose to neutrals due to elastic and inelastic interactions. The portrayed quantities have been evaluated by means of \eqref{eq:elasticLoss}, the subscript refers to the particular interaction: ion-neutral charge exchange (\emph{cex}), ion-neutral momentum exchange (\emph{mex}), electron-neutral ionisation (\emph{ion}), electron-neutral lumped excitation 1S and 2P (\emph{1S} and \emph{2P} respectively), and electron-neutral elastic scattering (\emph{scat}).
As expected, collision-related losses considerably increase while increasing the neutral number density up to $10^{18}$~m$^{-3}$.
Considering first the low power regime in Fig.~\ref{fig:powerBar}, it is quite evident how the two main interactions losses are the ion-neutral momentum exchange ($P_{mex}$) and charge exchange ($P_{cex}$). Other collisional terms are negligible due to their low frequency. The main reason is that the average energy possessed by the electrons produced in a low operating power scenario is not enough to consistently cause excitation and ionisation phenomena. Only the electron population of the high velocity tail of the Maxwellian distribution is able to trigger those interactions, causing the overall power contribution to be negligible.

A different scenario is presented for the 130 W case where ionisation interaction becomes the main cause of energy loss. Excitation processes are non-negligible as well. For the lower background pressure cases, charge and momentum exchange phenomena seem to be negligible, becoming of relevant importance only for $P_{a}>10^{-3}$, where their power losses are quite close to the excitation ones.

Taking a more general look to the plot reveals how the presence of a background neutral gas density highly affects the total power loss caused by inter-species interactions, starting from a mere 2\% of the total power delivered to the plume up to approximately 10\% and 22\% for the low and high power cases respectively.\\ 

\subsection{Thrust comparison and limiting assumptions}
\begin{figure}[!h]
    \centering
    \includegraphics[width=0.7\linewidth]{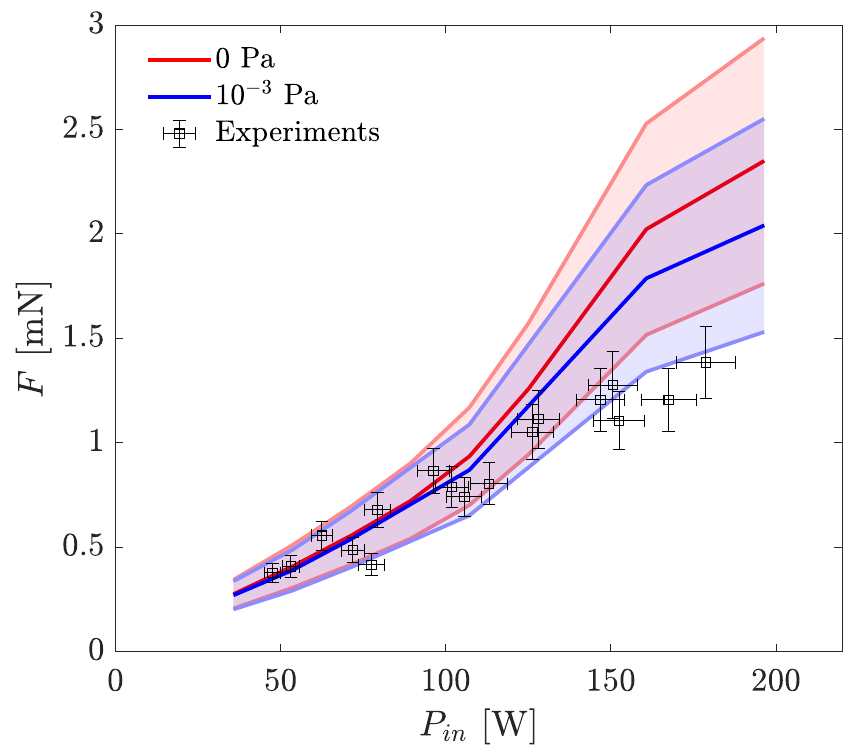}
    \caption{Thrust evaluated for different pressure levels against experimental measures.}
    \label{fig:thrustExp}
\end{figure}
To test the validity of the findings, this section will compare the numerical results to experimental measures of thrust that have been made at finite background pressure. 
Measurements have been performed at the High Vacuum Facilities of the University of Padova; a vacuum chamber of radius 0.3 m and length 2 m, maintained at a working pressure of circa $10^{-3}$ Pa. A detailed description of the experimental setup can be found in reference \cite{b:bellomo2021}. The thrust balance is described in \cite{b:trezzolani}; measurements error are estimated to be approximately 25\% for thrust and 10\% for the operating power \cite{b:trezzolani}. 
At the same time, the numerical error bands arise from different factors, such as assumptions in the GSM and uncertainty in the radio-frequency efficiency, as well as the uncertainty in collision cross sections used both in the GSM and in the PIC \cite{b:souhairPP}, and statistical variance in the PIC method \cite{b:andrews2022}. The error bands are approximately of 25\% of the simulated thrust. 
Fig.~\ref{fig:thrustExp} presents the estimated thrust values from the numerical suite for 0 and $1\times 10^{-3}$ Pa against the experimental measures. For the ideal vacuum simulations, there is good agreement between $P_{in}=50-120$ W; the experimental measure lie within the 25\% error of the numerical model. However, at higher values of $P_{in}$ it is clear that the numerical routine increasingly overestimates the thrust, with an error of +59\% at 180 W. As already seen, the introduction of a background neutral \textcolor{black}{population}, that replicates the facility environment, decreases the thrust, reducing the overestimation.
The plateau of experimental thrust at increasing power (and increasing $T_e$) might well be explained by losses in MN efficiency due to the chamber pressure. 
The agreement with experiments is considered acceptable ($<30\%$) since the numerical model is affected by several assumptions.
For example, the effects of background density on the ionisation process of the source region have been neglected. In a real scenario, the overall source ion-neutral ratio might change with increasing pressure, contributing to the thruster performance alteration.

The numerical acceleration methods might introduce some additional error, however, a study conducted by Andrews et al. \cite{b:andrews2022} has shown how the thrust estimation is basically unaffected (variation lower than $1\%$) by $\gamma$ and $f$ as long as their value remains limited to a reasonable range ($\gamma < 50$ and $f = 250$).

In the PIC, the neutral gas field is included with a uniform distribution. In practice, the position of the vacuum pumps and material sputtering would affect the local background density. It was seen in Fig.~\ref{fig:thrustLoss} that significant thrust losses can result from even small increases in pressure beyond $1\cdot 10^{-3}$ Pa, and there may be a significant uncertainty in the assumed working pressure at the MN throat region.

\section{Conclusion}
The main goal of this work was the investigation of the facilities effects on the performance of a cathode-less RF thruster. Several levels of background pressures have been tested for four different operational regimes: low power 60 W and 90 W and high power 130 W and 150 W cases were considered for the study.\\
The additional background neutral population caused a degradation of the thruster performance due to the decisive increase of the electron-neutral and ion-neutral elastic and inelastic interactions. More in detail, for low operating power (i.e. low plasma temperature), the average energy possessed by electrons was too low to generate ionisation and excitation interactions in a consistent way, as a consequence the main factors of power loss in that case were proven to be ion-neutral charge and momentum exchange. Adversely, in the 130-150 W cases, electron-based inelastic collision dominated the plume causing up the 70\% of the total power losses.\\
These enhanced collision rates resulted in an overall thrust decrease, with a more decisive impact on the high-power scenario, where a drop of more than 13\% of the exerted force was observed. The low power case presented a decreasing trend in the performance as well, however, the thrust depletion was limited to less than 2\% even at the highest background pressure level.
This effect is related to the higher neutral particle count that is ejected by the source when operating the engine at low power, causing the plume to have a similar or even higher neutral density than the background for the first portion of the domain. 

In conclusion facilities effects definitely represent an issue in the characterisation of RF source-based MEPTs: once the plasma temperature of the plume becomes high enough (say $>$8 eV for xenon) to cause ionisation collisions in a consistent manner, performing measurements for high-power level regimes at a pressure circa 10$^{-3}$ Pa, or higher, may yield less accurate results. Despite being less affected by the ambient neutral count, also low power thrusters (i.e. characterised by $T_e<8$~eV) shall be kept at a pressure lower than $10^{-3}$ Pa, in order to avoid undesired facility-related performance drop due to ion-neutral collisions.

\section*{Acknowledgements}
We acknowledge Technology for Propulsion and Innovation (T4i) S.p.A. for the support provided in the development of this work. We also acknowledge the CINECA award under the ISCRA initiative, for the availability of high-performance computing resources and support.

\appendix
\section{Global Model}
\label{sec:appendixA}
\setcounter{section}{1}
In this appendix, the expressions for the evaluation of \eqref{eqn:Gm_mass} and \eqref{eqn:Gm_power} have been reported. For a more detailed description of the model as a whole we address the reader to \cite{b:bosi2019,b:lieberman2005}.
Starting with \eqref{eqn:Gm_mass}, 
\begin{equation}
    R_{chem}^I = \sum_J K_{JI}n_Jn_e - \sum_J K_{IJ}n_In_e 
\end{equation}
is the species density flux associated to chemical reactions. $K_{IJ}$ is the rate constant for inelastic transition from species $I$ to $J$. 
The term 
\begin{equation}
    R_{wall}^I = \frac{S_{wall}^I}{\mathcal{V}}\Gamma_{wall}^I
\end{equation}
is related to wall production and losses. Here, $\mathcal{V} = \mathcal{L}\pi\mathcal{R}^2$ represents the volume of the source ($\mathcal{R}$ and $\mathcal{L}$ being the source radius and length, respectively), $\Gamma_{wall}^I = n_Iu_B$ is the particle flux, $u_B$ being the Bohm speed. Lastly, $S_{wall}^I$ is an equivalent source (or well) surface; more in detail, for a closed cylinder with cusps, the surface calculation reads
\begin{equation}\label{eqn:wallFlux}
    S_{wall}^I = 2\pi\mathcal{R}h_L\beta + h_{R\perp}(2\pi\mathcal{R}\mathcal{L} - S_{cusps} + h_{R\parallel})S_{cusps}
\end{equation}
Despite the assumption of uniform magnetostatic field, the presence of cusps is included by means of an empirical model. The evaluation of the cusp area can hence be performed as shown by \emph{Goebel et al.} \cite{b:goeb2008}
\begin{equation}
    S_{cusps} = 4N_{cusps}\sqrt{r_{ce}r_{ci}}2\pi\mathcal{R}
\end{equation}
within these two expressions, $h_L, h_R$ and $\beta$ are semi-empirical coefficients that take into account the non-uniformity of the plasma profiles \cite{b:lee1995}. 
$N_{cusps}$ is the number of magnetic cusps, while $r_{ci}$ and $r_{ce}$ are the ion and electron cyclotron radii, respectively.

The last term of \eqref{eqn:Gm_mass} is the flux density at the source outlet,
\begin{equation}
    R_{ex}^I = \frac{S_{ex}^I}{\mathcal{V}}\Gamma_{ex}^I
\end{equation}
$S_{ex}^I$ is the physical area of the exit section; $\Gamma^e = \Gamma^i = n_eu_B$ for ions and electrons while the neutral flux is computed assuming free molecular regime $\Gamma^g = 1/4n_gu_{th}$ ($u_{th}$ being the neutrals thermal speed).

Also in the case of the electron power equation \eqref{eqn:Gm_power}, a chemical contribution can be highlighted, 
\begin{equation}
    P_{chem} = \sum_I\sum_J K_{IJ}n_In_e\Delta U_{IJ} + \sum_I K_{II}n_In_e\frac{3m_e}{m_I}T_e
\end{equation}
where $\Delta U_{IJ}$ represents the energy difference between between species $I$ and $J$ in eV.
Regarding the elastic collisions term, $K_{II}$ is the rate constant for the interaction between species $I$ and electrons, while $m_I$ refers to the species $I$ mass.
The well/source term for the wall-plasma interaction can be evaluated as \cite{b:Ahedo2013,b:lafleur2014}
\begin{equation}\label{eqn:wallPower}
    P_{wall} = R_{wall}^e\left(\frac{5}{2} + \frac{1}{2}\log\sqrt{\frac{m_i}{2\pi m_e}} \right)T_e
\end{equation}
note that for expressions \eqref{eqn:wallFlux} and \eqref{eqn:wallPower} to be valid, the Bohm sheath criterion at the source wall must be assumed \cite{b:guaita2022}.
Similarly, the exhaust power contributions $P_{ex}$ reads,
\begin{equation}
    P_{ex} = R_{ex}^e\left(\frac{5}{2} + \frac{1}{2}\log\sqrt{\frac{m_i}{2\pi m_e}} \right) T_e
\end{equation}
The computation of the reaction rate coefficient $K_{IJ}$ is carried out through
\begin{equation}
    K = \sqrt{\frac{2q}{m_e}}\int_0^\infty\varepsilon\sigma f_0 d\varepsilon
\end{equation}
$q$ and $\varepsilon$ being the electron charge and energy (in eV), respectively. $\sigma$ is the collision cross-section for the generic particle reaction, whereas $f_0$ is the electron energy distribution function (EEDF). Within this study, the hypothesis of Maxwellian distribution is made \cite{b:hagelaar2005}
\begin{equation}
    f_0(\varepsilon) = 2\sqrt{\frac{1}{T_e^3\pi}}\exp\left(-\frac{\varepsilon}{T_e}\right)
\end{equation}
\clearpage
\bibliographystyle{elsarticle-num} 
\bibliography{cas-refs}





\end{document}